\documentclass[12pt]{article}
\usepackage{amssymb,amsmath}

\newcommand{\IR}{\mathbb{R}}
\def\be{\begin{equation}}
\def\bea{\begin{eqnarray}}
\def\ee{\end{equation}}
\def\eea{\end{eqnarray}}
\def\nn{\nonumber \\}

\begin{document}
\begin{titlepage}

\begin{flushright}
IPhT-T09/113
\end{flushright}

\bigskip
\bigskip
\centerline{\Large \bf Multi-Center non-BPS Black Holes - the Solution}
\bigskip
\bigskip
\centerline{{\bf Iosif Bena$^1$, Stefano Giusto$^1$,}}
\centerline{{\bf Cl\'{e}ment Ruef$^{\, 1}$ and Nicholas P. Warner$^2$}}
\bigskip
\centerline{$^1$ Institut de Physique Th\'eorique, }
\centerline{CEA Saclay, 91191 Gif sur Yvette, France}
\bigskip
\centerline{$^2$ Department of Physics and Astronomy}
\centerline{University of Southern California} \centerline{Los
Angeles, CA 90089, USA}
\bigskip
\centerline{{\rm iosif.bena@cea.fr,~stefano.giusto@cea.fr,~clement.ruef@cea.fr,~warner@usc.edu } }
\bigskip
\bigskip

\begin{abstract}

\noindent We construct multi-center, non-supersymmetric four-dimensional
solutions describing a rotating $\overline{\rm D6}$-D2 black hole and an arbitrary number of
D4-D2-D0 black holes in a line. These solutions correspond
to an arbitrary number of extremal non-BPS black rings in a Taub-NUT
space with a rotating three-charge black hole in the middle. The
positions of the centers are determined by solving a set of
``bubble'' or ``integrability'' equations that contain cubic
polynomials of the inter-center distance, and that allow scaling
solutions even when the total four-dimensional angular momentum of
the scaling centers is non-zero.

\end{abstract}

\end{titlepage}

%%%%%%%%%%%%%%%%%%%%%%%%%%%%%%%%%%%%%
\section{Introduction}
%%%%%%%%%%%%%%%%%%%%%%%%%%%%%%%%%%%%%

Multi-center BPS black-hole solutions in four dimensions, and their
five-dimensional counterparts,
\cite{denef,Gauntlett:2002nw,Gauntlett:2004qy,Bena:2005va,Berglund:2005vb,Saxena:2005uk}
have played a crucial role in several ares of research aimed at
understanding the quantum structure of black holes in string theory.
These areas include the relation between five-dimensional black rings
and four-dimensional black holes \cite{4D5Dring}, the ``proof'' and
``disproof'' \cite{denef-moore} of the OSV conjecture \cite{OSV},
the construction of smooth horizonless solutions that describe black-hole microstates in the same regime of parameters where the
classical black hole exist \cite{us,Bena:2007qc}, the construction of entropy enigmas
\cite{denef-moore}, the calculation of index-jumps when crossing walls
of marginal stability \cite{stability}, and the realization that
quantum effects can wipe away a
macroscopic region of a smooth horizonless low-curvature solution
\cite{Bena:2007qc,sheer}.

Given the large amount of knowledge about BPS black holes that has
been obtained by studying multi-center solutions, it is natural to
ask whether these solutions can be generalized to non-BPS black holes.
This problem appears, {\it a priori,} rather hopeless, given that one is
looking for four- or five-dimensional non-supersymmetric solutions of
Einstein's equations that depend on at least two variables, and that
these equations generically do not ``factorize'' into first-order equation
(as they do for BPS systems \cite{Bena:2004de,Gutowski:2004yv}). Indeed, most of the known
solutions have been constructed in an ``artisanal'' fashion, and are
either essentially two-centered
\cite{JMaRT-all}
or have no $E \times B$ interactions between the centers
\cite{Gaiotto-Finn}\footnote{Hence, these configurations are more similar
  in spirit to Majumdar-Papapetrou multi-center solutions than to the
  ones of \cite{denef}}.

The best target for a systematic construction of multi-center, non-BPS
black holes are extremal solutions. Indeed, for single-center configurations the equations
underlying these solutions have been shown to factorize
\cite{Ceresole:2007wx,factorization}. Furthermore, Goldstein and
Katmadas have observed \cite{Goldstein} that one can construct
a specific class of ``almost-BPS'' solutions by solving the same linear system of
equations as for BPS solutions but on a four-dimensional base space of
reverse orientation. This observation has led to the explicit
construction, in \cite{BDGRW}, of the seed solution for the most general
rotating extremal black hole in ${\cal N}=8$ supergravity in four
dimension, and of a solution describing a non-BPS black ring in
Taub-NUT. This latter solution descends into four dimensions as a
two-centered solution in which one of the centers is a rotating  $\overline{\rm D6}$-D2
black hole, and the other center is a D4-D2-D0 black hole.

Our purpose in this paper is to extend this construction and build non-BPS
solutions that contain a black hole and an arbitrary number of
concentric black rings in Taub-NUT.  As in \cite{BDGRW}, these
solutions have a non-trivial four-dimensional angular momentum\ that
comes both from the rotation of the black hole and from the $E \times
B$ interactions between the black-hole and the black-ring centers, and
between black-ring and black-ring centers. Hence, for generic charges
our solution can be described  in terms of a quiver that has
arrows running between every pair of points.

Just as for BPS multi-center solutions, the locations of the centers
are not arbitrary: the absence of closed time-like curves and of Dirac
strings imposes certain ``bubble'' or ``integrability'' equations that
these locations must satisfy. However, unlike the BPS bubble
equations, that are linear in the inverse of the inter-center
distances, the non-BPS bubble equations have denominators that are
cubic polynomials in the inter-center distances. Moreover, both the
two-centered and the multi-centered solutions have walls of marginal
stability in the moduli space, across which the solutions can disappear.

Another important aspect of the non-BPS bubble equations is that they
admit scaling solutions. Furthermore, when one of the scaling centers
is the rotating black hole at the center of Taub-NUT, the total
four-dimensional angular momentum of the scaling centers can remain
large throughout the scaling! The throat of the non-BPS scaling
solutions then asymptotes to the (intrinsically non-BPS) throat of a
rotating four-dimensional black hole. This makes our scaling solutions more
general than the BPS ones (whose four-dimensional angular
momentum always goes to zero in the scaling limit).

Interestingly enough, in the scaling regime, the non-BPS bubble
equations equations become identical to the BPS bubble equations. For
scaling solutions with vanishing four-dimensional angular momentum, this is
to be expected: Indeed, as observed in \cite{Goldstein,BDGRW}, when
the Taub-NUT base space degenerates to $\IR^4$ or $\IR^3\times S^1$,
the almost-BPS solutions become identical to the BPS
ones\footnote{This fact was used extensively in \cite{BDGRW} and will
  also be used here to obtain solutions to the almost-BPS equations by
  recycling pieces of the BPS solution.}.  When the centers are very
close to each other, the harmonic function that determines the
Taub-NUT base can be approximated either by $1/r$ or by a constant
(depending on whether the Taub-NUT center is included in the scaling
solution or not).  Hence, as the centers scale they see a base space
that resembles with increasing accuracy the base of a BPS scaling
solution, and the non-BPS bubble equations asymptote to the BPS bubble
equations.  Putting it another way, the throat of a non-BPS, non-rotating
scaling solution in which the Taub-NUT center participates
increasingly resembles the throat a D2-D2-D2-$\rm \overline{D6}$
extremal non-BPS single-center black hole, which is the same as the
throat of its D2-D2-D2-D6 BPS cousin \cite{emparan-horowitz}. However,
it is rather mysterious why, in the presence of four-dimensional
angular momentum, the scaling limit of the non-BPS bubble equations is
still the same as that of the BPS ones, or, equivalently, why the addition
of four-dimensional angular momentum to the black hole center does not
affect the non-BPS bubble equations.

In Section 2 we find the metric warp factors, the electric and
magnetic field strengths, as well as the angular momentum vector of
our multi-center solutions.  In Section 3 we study the regularity
conditions imposed by the absence of closed time-like curves (CTC's), and find
the ``bubble'' or ``integrability'' equations that the positions of
the centers must satisfy. We also study regularity at the black-hole
and black-ring horizons, and relate the charges that appear in the
supergravity solution to quantized charges. We conclude this section
by investigating scaling solutions. We present conclusions and potential future
directions of research in Section 4.

%%%%%%%%%%%%%%%%%%%%%%%%%%%%%%%%%%%%%
\section{Multi-center non-BPS solutions in Taub-NUT}
%%%%%%%%%%%%%%%%%%%%%%%%%%%%%%%%%%%%%

%%%%%%%%%%%%%%%%%%%%%%%%%%%%%
\subsection{The Ansatz and the almost-BPS equations}
%%%%%%%%%%%%%%%%%%%%%%%%%%%%%

As observed in \cite{Goldstein,BDGRW}, both BPS and almost BPS  solutions of eleven-dimensional supergravity carrying M2 and M5 charges are of the form:

\bea
ds^2&\!\!\!\!=\!\!\!\!& -(Z_1 Z_2 Z_3 )^{-2/ 3}(dt+k)^2 + (Z_1 Z_2 Z_3)^{1/3} ds^2_4 \nn
&&+ \Bigl({Z_2 Z_3\over Z_1^2}\Bigr)^{1/3} (dx_1^2+dx_2^2)+\Bigl({Z_1 Z_3\over Z_2^2}\Bigr)^{1/3} (dx_3^2+dx_4^2)+\Bigl({Z_1 Z_2\over Z_3^2}\Bigr)^{1/3} (dx_5^2+dx_6^2)\label{11D}\\
C^{(3)}&\!\!\!\!=\!\!\!\!&\! \Bigl(a^1 - {dt+k\over Z_1}\Bigr)\!\!\wedge \!dx_1\!\wedge \!dx_2 +\!  \Bigl(a^2 - {dt+k\over Z_2}\Bigr)\!\!\wedge\! dx_3\!\wedge\! dx_4 +\! \Bigl(a^3 - {dt+k\over Z_3}\Bigr)\!\!\wedge\! dx_5\!\wedge\! dx_6 \,,
\eea
where $ds^2_4$ is a hyper-K\"ahler, four-dimensional metric whose curvature we take to be self-dual.

The almost-BPS solutions are given by:
\bea
&&\Theta^{(I)} =- *_4 \Theta^{(I)} \label{thetaeq}\\
&&d*_4d Z_I = {C_{IJK}\over 2} \Theta^{(I)} \wedge \Theta^{(I)}\label{zeq}\\
&&dk-*_4 dk = Z_I  \Theta^{(I)}\label{keq}\,,
\eea
where $*_4$ is the Hodge duality operator for the metric $ds^2_4$, and
the anti-self-dual dipole field strengths are defined as $\Theta^{(I)}\equiv d a^I$.
Note that if one considers these equations on a hyper-K\"ahler base
with an anti-self-dual curvature, they describe BPS
solutions.  Supersymmetry is broken
only because the curvature of the base and the two-form dipole field strengths have opposite
orientations.

%%%%%%%%%%%%%%%%%%%%%%%%%%%%%
\subsection{Solutions with a Taub-NUT base}
%%%%%%%%%%%%%%%%%%%%%%%%%%%%%

Our purpose is to construct multi-center solutions with a Taub-NUT base:
\be
ds^2_4 = V^{-1}(d\psi+A)+V ds^2_3
\ee
with
\be
V= h + {q\over r}\,,\quad A= q \cos\theta d\phi\,,\quad ds^2_3 = dr^2 + r^2 d\theta^2+r^2\sin^2\theta d\phi^2\,.
\ee
Let $a_i$, $i=1,\dots,N$ denote a succession of points along the $z$ axis in $\mathbb{R}^3$, distinct from the Taub-NUT origin ($a_i\not =0$). In the  $\mathbb{R}^3$ base of the Taub-NUT space, the distance from a given point $(r,\theta,\phi)$ to any one of these points is
\be
\Sigma_i = \sqrt{r^2+a_i^2 -2 r a_i \cos\theta} \,,
\ee
and the polar angle of that point with respect to the point $i$ is
\be
\cos\theta_i = {r\cos\theta-a_i\over \Sigma_i}\,.
\ee

As shown in \cite{Goldstein,BDGRW}, the  M5 (magnetic) charges are determined by harmonic functions $K^{(I)}$, and we assume that they have generic poles at the points $a_i$\footnote{Allowing  $K^{(I)}$ to have poles at $r=0$ appears to lead to singular solutions.}
\be
K^{(I)}=\sum_{i=1}^N {d^{(I)}_i\over \Sigma_i}\,.
\label{K}
\ee
The harmonic functions $L_I$ associated with the M2 (electric) charges can have poles both at the points, $a_i$, and at the Taub-NUT center:
\be
 L_I  ~=~ \ell_I  + {Q^{(I)}_0\over r}+ \sum_i {Q^{(I)}_i\over \Sigma_i}~=~   \ell_I + \sum_{i=0}^N {Q^{(I)}_i\over \Sigma_i} \,,
 \label{L}
\ee
where  $\Sigma_0 \equiv r$. A solution of the almost-BPS equations (\ref{thetaeq}), (\ref{zeq}) and (\ref{keq}) can now be constructed from these harmonic functions.

%%%%%%%%%%%%%%%%%%%%%%%%%%%%%
\subsection{Dipole field strengths}
%%%%%%%%%%%%%%%%%%%%%%%%%%%%%

The two-form field strengths, $\Theta^{(I)}$,  are closed and anti-self dual in the Taub-NUT space and have the form:
\be
\Theta^{(I)} = d[K^{(I)} (d\psi+A)+b^{(I)}]\,,
\label{theta}
\ee
where $K^{(I)}$ is given in (\ref{K}) and $b^{(I)}$ is given by
\bea
*_3 d b^{(I)} = V d K^{(I)} - K^{(I)} dV \,\,\Rightarrow\,\, b^{(I)}=\sum_i  {d^{(I)}_i\over \Sigma_i} \Big(h (r\cos\theta-a_i)+q \,{r-a_i \cos\theta\over a_i}\Big) d\phi\,.
\eea

%%%%%%%%%%%%%%%%%%%%%%%%%%%%%
\subsection{Warp factors}
%%%%%%%%%%%%%%%%%%%%%%%%%%%%%

The warp factors, $Z_I$, which encode the M2 charges, are determined by (\ref{zeq}), and for
 the dipole field strengths in (\ref{theta}) this equation becomes:
\bea
\Box_3 Z_I =  V {|\epsilon_{IJK}|\over 2}\Box_3 (K^{(J)} K^{(K)}) = \Bigl(h+{q\over r}\Bigr) \sum_{j,k} {|\epsilon_{IJK}|\over 2}\Box_3 \Bigl({d^{(J)}_j d^{(K)}_k \over \Sigma_j \Sigma_k}\Bigr)\,,
\label{warpeq}
\eea
where sums over repeated $J,K$ indices are implicit (as they will be throughout this paper). It is completely trivial to solve this equation for the terms proportional to $h$ and for the term proportional to $q$ we use the identity:
\be
\Box_3\Bigl[ {r\over a_i a_j} {1\over \Sigma_i \Sigma_j} \Bigr]= {1\over r} \Box_3\Bigl[{1\over \Sigma_i \Sigma_j}\Bigr]\,.
\ee
If one also includes the freedom to add to $Z_I$ a generic harmonic function, $L_I$, given in (\ref{L}), the complete solution for $Z_I$ is
\bea
Z_I =~=~ L_I + {|\epsilon_{IJK}|\over 2}\sum_{j,k}
\Bigl(h+{q r\over a_j a_k}\Bigr) {d^{(J)}_j d^{(K)}_k \over \Sigma_j \Sigma_k}\,.
\eea

%%%%%%%%%%%%%%%%%%%%%%%%%%%%%
\subsection{The angular momentum one-form}
%%%%%%%%%%%%%%%%%%%%%%%%%%%%%

The angular momentum one-form, $k$, can be decomposed as
\be
k=\mu(d\psi+A)+\omega\,,
\ee
where $\omega$ is a one-form on $\IR^3$.
Equation (\ref{keq}) then becomes\footnote{All sums over $i,i',j,k$ are from $0$ to $N$, with the convention that $d^{(I)}_0=0$.}:
\bea
&&d(V\mu)+ *_3 d\omega = V Z_I  d K^{(I)} \nonumber\\
&& =V  \sum_i \ell_I d^{(I)}_i d{1\over \Sigma_i} +\Bigl(h+{q\over r}\Bigr) \sum_{i,i'} Q^{(I)}_i d^{(I)}_{i'} {1\over \Sigma_i}d{1\over \Sigma_{i'}}\nonumber\\
&&+  {|\epsilon_{IJK}|\over 2} \sum_{i,j,k} d^{(I)}_i d^{(J)}_j d^{(K)}_k  \Bigl[h^2+{q^2\over a_j a_k} + h q \Bigl({1\over r}+{r\over a_j a_k}\Bigr)\Bigr] {1\over \Sigma_j\Sigma_k} d{1\over \Sigma_i}  \,.
\eea
It is convenient to rewrite the term cubic in $d^{(I)}_i$ as
\bea
&& {|\epsilon_{IJK}|\over 2} \sum_{i,j,k} d^{(I)}_i d^{(J)}_j d^{(K)}_k  \Bigl[h^2+{q^2\over a_j a_k} + h q \Bigl({1\over r}+{r\over a_j a_k}\Bigr)\Bigr] {1\over \Sigma_j\Sigma_k} d{1\over \Sigma_i} \nonumber\\
&&\qquad=\sum_{i,j,k}  d^{(1)}_i d^{(2)}_j d^{(3)}_k (h^2\, T^{(1)}_{ijk} + q^2 \,T^{(2)}_{ijk} + h q \,T^{(3)}_{ijk})\,,
\label{gensrc}
\eea
where
\bea
T^{(1)}_{ijk}&\equiv& {1\over \Sigma_j\Sigma_k} d{1\over \Sigma_i} + {1\over \Sigma_i\Sigma_k} d{1\over \Sigma_j} + {1\over \Sigma_i\Sigma_j} d{1\over \Sigma_k} \nonumber\\
T^{(2)}_{ijk}&\equiv& {1\over a_j a_k} {1\over \Sigma_j\Sigma_k} d{1\over \Sigma_i} +{1\over a_i a_k} {1\over \Sigma_i\Sigma_k} d{1\over \Sigma_j} +{1\over a_i a_j} {1\over \Sigma_i\Sigma_j} d{1\over \Sigma_k} \nonumber\\
T^{(3)}_{ijk}&\equiv& \Bigl({1\over r}+{r\over a_j a_k}\Bigr) {1\over \Sigma_j\Sigma_k} d{1\over \Sigma_i} + \Bigl({1\over r}+{r\over a_i a_k}\Bigr) {1\over \Sigma_i\Sigma_k} d{1\over \Sigma_j} +  \Bigl({1\over r}+{r\over a_i a_j}\Bigr){1\over \Sigma_i\Sigma_j} d{1\over \Sigma_k} \,,
\label{symmsrcs}
\eea
with $a_i$, $a_j$, $a_k$ any three, possibly coincident, non-vanishing points.  Note that in (\ref{symmsrcs}) we have explicitly symmetrized over the three source points and so there is an associated factor of $1/3$ but this is canceled in  (\ref{gensrc}) by the explicit replacement of ${1 \over 2}{|\epsilon_{IJK}|}$.

One can thus reduce the complete solution for $\mu$ and $\omega$ to the solution of the following equations:
\bea
&&d(V\mu^{(1)}_{i})+ *_3 d\omega^{(1)}_{i}=V d {1\over \Sigma_i}\nonumber\\
&&d(V\mu^{(2)}_{i})+ *_3 d\omega^{(2)}_{i}= {1\over \Sigma_i} d{1\over \Sigma_i}\quad (i\not= 0)\nonumber\\
&&d(V\mu^{(3)}_{i j})+ *_3 d\omega^{(3)}_{ i j}={1\over \Sigma_i} d{1\over \Sigma_j}\quad (i\not = j)\nonumber\\
&&d(V\mu^{(4)}_{i})+ *_3 d\omega^{(4)}_{ i}={1\over r\,\Sigma_i} d{1\over \Sigma_i}\quad (i\not =0)\nonumber\\
\eea
\bea
&&d(V\mu^{(5)}_{i j})+ *_3 d\omega^{(5)}_{ i j}={1\over r\, \Sigma_i} d{1\over \Sigma_j}\quad (i\not = j ,j\not =0 )\nonumber\\
&&d(V\mu^{(6)}_{i j k})+ *_3 d\omega^{(6)}_{ i j k }= T^{(1)}_{i j k}\quad (i,j,k\not = 0)\nonumber\\
&&d(V\mu^{(7)}_{i j k})+ *_3 d\omega^{(7)}_{ i j k }= T^{(2)}_{i j k}\quad (i,j,k\not = 0)\nonumber\\
&&d(V\mu^{(8)}_{i j k})+ *_3 d\omega^{(8)}_{ i j k }= T^{(3)}_{i j k}\quad (i,j,k\not = 0)\,.
\eea

A solution to this is:
\bea
&&V\mu^{(1)}_i ={V \over 2 \Sigma_i}\,,\quad \omega^{(1)}_i = {h\over 2}{r\cos\theta-a_i\over \Sigma_i}d\phi+{q\over 2} {r-a_i\cos\theta\over a_i \Sigma_i}d\phi\nonumber\\
&&V\mu^{(2)}_i = {1\over 2 \Sigma^2_i}\,,\quad \omega^{(2)}_i =0\nonumber\\
&&V\mu^{(3)}_{ij} = {1\over 2}{1\over \Sigma_i\Sigma_j}\quad  \omega^{(3)}_{ij}={r^2 + a_i a_j -  (a_i+a_j) r \cos\theta\over 2 (a_j-a_i) \Sigma_i\Sigma_j}d\phi\nonumber\\
&&V\mu^{(4)}_i = {\cos\theta\over 2 a_i \Sigma_i^2}\,,\quad \omega^{(4)}_i={r\sin^2\theta\over 2 a_i \Sigma_i^2} d\phi\nonumber\\
&&V\mu^{(5)}_{ij}={r^2 + a_i a_j - 2 a_j r \cos\theta \over 2 a_j (a_i-a_j) r \Sigma_i \Sigma_j}\,,
\quad \omega^{(5)}_{ij} = {r(a_i+a_j \cos2\theta)-(r^2 + a_i a_j)\cos\theta\over 2 a_j (a_i-a_j)\Sigma_i \Sigma_j}d\phi\nonumber\\
&& V \mu^{(6)}_{ijk} = {1\over \Sigma_i \Sigma_j \Sigma_k}\,,\quad \omega^{(6)}_{ijk}=0\nonumber\\
&& V \mu^{(7)}_{ijk} ={r\cos\theta\over a_i a_j a_k \Sigma_i \Sigma_j \Sigma_k}\ \,,\quad \omega^{(7)}_{ijk}= {r^2 \sin^2\theta\over a_i a_j a_k \Sigma_i \Sigma_j \Sigma_k} d\phi\nonumber\\
&&V \mu^{(8)}_{ijk}={r^2 (a_i + a_j + a_k)+ a_i a_j  a_k\over 2 a_i a_j a_k r \Sigma_i \Sigma_j \Sigma_k}\nonumber\\
&&\qquad \omega^{(8)}_{ijk}={r^3 +r(a_i a_j+ a_i a_k + a_j a_k)-(r^2(a_i+a_j+a_k)+a_i a_j a_k) \cos\theta\over 2 a_i a_j a_k \Sigma_i \Sigma_j\Sigma_k}d\phi\,.
\eea

One can also add to $k$ a solution of the homogeneous equation $dk-*_4 dk=0$, and we consider a such solution with components:
\be
V \mu^{(9)} = M \,,\quad *_3 d\omega^{(9)}= - dM\,,
\ee
where $M$ is a harmonic function that generically can be of the form:
\be
M = m + \sum_{i=0}^N {m_i\over \Sigma_i}+\sum_{i=0}^N \alpha_i {\cos\theta_i\over \Sigma_i^2}\,.
\ee
Note that we have allowed for the possibility of dipole harmonic
functions in $M$ because we know, from the two-center solution
\cite{BDGRW}, that these are necessary to obtain a rotating black
hole at the Taub-NUT center. The corresponding $\omega^{(9)}$ is:
\be
 \omega^{(9)}=\kappa d\phi - \sum_{i=0}^N m_i \cos\theta_i d\phi +\sum_{i=0}^N \alpha_i {r^2\sin^2\theta\over \Sigma_i^3}d\phi\,.
\ee
The complete expression for $\mu$ and $\omega$ is then
\bea
&&\mu = \sum_{i} \ell_I d^{(I)}_i \mu^{(1)}_i + \sum_{i} Q^{(I)}_i d^{(I)}_i (h \mu^{(2)}_i + q \mu^{(4)}_i)+  \sum_{i\not =i'} Q^{(I)}_i d^{(I)}_{i'} (h \mu^{(3)}_{ii'} + q \mu^{(5)}_{ii'})\nonumber\\
&&\qquad +
\sum_{i,j,k} d^{(1)}_i d^{(2)}_j d^{(3)}_k (h^2 \mu^{(6)}_{ijk}+ q^2  \mu^{(7)}_{ijk}+ h q  \mu^{(8)}_{ijk}) +\mu^{(9)}\label{fullmu} \\
&&\omega = \sum_i \ell_I  d^{(I)}_i \omega^{(1)}_i + \sum_i Q^{(I)}_i d^{(I)}_i (h \omega^{(2)}_i + q \omega^{(4)}_i)+  \sum_{i\not =i'} Q^{(I)}_i d^{(I)}_{i'} (h \omega^{(3)}_{ii'} + q \omega^{(5)}_{ii'})\nonumber\\
&&\qquad +
\sum_{i,j,k} d^{(1)}_i d^{(2)}_j d^{(3)}_k (h^2 \omega^{(6)}_{ijk}+ q^2  \omega^{(7)}_{ijk}+ h q  \omega^{(8)}_{ijk}) +\omega^{(9)}\,,
\label{fullomega}
\eea
or, more explicitly,
\bea
&&\mu = \sum_{i} {\ell_I d^{(I)}_i\over 2 \Sigma_i}+\sum_{i} {Q^{(I)}_i d^{(I)}_i\over 2 V \Sigma_i^2}\Bigl(h+{q\cos\theta\over a_i}\Bigr)+\sum_{i\not=i'}{Q^{(I)}_i d^{(I)}_{i'}\over 2 V \Sigma_i\Sigma_{i'}} \Bigl(h+q {r^2 + a_i a_{i'} - 2 a_{i'} r \cos\theta\over a_{i'} (a_i - a_{i'})}\Bigr)\nonumber\\
&&\qquad+\sum_{i,j,k} {d^{(1)}_i d^{(2)}_j d^{(3)}_k\over V \Sigma_i\Sigma_J\Sigma_k} \Bigl(h^2+q^2{r\cos\theta\over a_i a_j a_k}+h q {r^2 (a_i + a_j + a_k)+ a_i a_j  a_k\over 2 a_i a_j a_k r }\Bigr)+{M\over V}\,,
\label{fullmu-exp}\\
&&\omega= \sum_{i} { \ell_I  d^{(I)}_i\over 2 \Sigma_i}\Bigl(h(r\cos\theta-a_i)+q {r-a_i \cos\theta\over a_i}\Bigr)d\phi+\sum_{i} Q^{(I)}_i d^{(I)}_i {q r \sin^2\theta\over 2 a_i \Sigma_i^2} d\phi\nonumber\\
&&\qquad+\sum_{i\not= i'} {Q^{(I)}_i d^{(I)}_{i'}\over 2 (a_{i'}-a_i)\Sigma_i \Sigma_{i'}}\Bigl(h (r^2 + a_i a_{i'} -  (a_i+a_{i'}) r \cos\theta)\nonumber\\
&&\hspace{7cm}- q{r(a_i+a_{i'} \cos2\theta)-(r^2 + a_i a_{i'})\cos\theta\over a_{i'} }\Bigr)d\phi\nonumber\\
&&\qquad+\sum_{i,j,k} {d^{(1)}_i d^{(2)}_j d^{(3)}_k\over a_i a_j a_k \Sigma_i\Sigma_J\Sigma_k} \Bigl(q^2 r^2 \sin^2\theta \nonumber\\
&&\hspace{4cm}+h q {r^3 +r(a_i a_j+ a_i a_k + a_j a_k)-(r^2(a_i+a_j+a_k)+a_i a_j a_k) \cos\theta\over 2 }\Bigr)d\phi\nonumber\\
&&\qquad +\kappa d\phi - \sum_{i=0}^N m_i \cos\theta_i d\phi +\sum_{i=0}^N \alpha_i {r^2\sin^2\theta\over \Sigma_i^3}d\phi \,.
\label{fullomega-exp}
\eea

%%%%%%%%%%%%%%%%%%%%%%%%%%%%%%%%%%%%%
\section{Regularity}
%%%%%%%%%%%%%%%%%%%%%%%%%%%%%%%%%%%%%

The solutions constructed above satisfy the equations of motion, but
are not necessarily regular. Indeed, the angular momentum one-form
$\omega$ is proportional to $d\phi$, and can have Dirac-Misner string
singularities, and these would lead to closed time-like curves (CTC's).  One must therefore require
$\omega$ to vanish on the $z$-axis (where the $\phi$
coordinate degenerates). Furthermore, near the poles of the
harmonic functions the warp factor and rotation one-form blow up, and
this can also lead to CTC's. We now find the
conditions that guarantee the absence of CTC's in these two
obvious places.

The conditions we will obtain are necessary but not
sufficient; to be absolutely sure of regularity and absence of CTC's one must usually
check each solution globally and in practice this is usually done individually and numerically.
Nevertheless, in our experience (and that of others \cite{gg1}), when
the charges and dipole charges of the rings have the same signs, and there are no
Dirac-Misner strings or CTC's at the horizons, the multi-center black ring solution is regular.

%%%%%%%%%%%%%%%%%%%%%%%%%%%%%
\subsection{Removing closed time-like curves}
%%%%%%%%%%%%%%%%%%%%%%%%%%%%%

We require $\omega_\phi$ to vanish for $\theta=0$ or $\pi$. Looking at
the various terms contributing to $\omega$ we see that only
$\omega^{(1)}$, $\omega^{(3)}$, $\omega^{(5)}$, $\omega^{(8)}$ and
$\omega^{(9)}$ are non-vanishing on the $z$-axis. Their values are:
\bea
&&\omega^{(1)}_i={s^{(-)}_i\over 2}\Bigl(h+{q\over a_i}\Bigr)d\phi\,,\quad \omega^{(3)}_{ij}={s^{(-)}_i s^{(-)}_j\over 2 (a_j -a_i)}d\phi\,, \quad \omega^{(5)}_{ij}={s^{(-)}_i s^{(-)}_j\over 2 a_j (a_j -a_i)}d\phi\,,\nonumber\\
&& \omega^{(8)}_{ijk}={s^{(-)}_i s^{(-)}_j s^{(-)}_k\over 2 a_i a_j a_k}d\phi\,,\quad \omega^{(9)}= (\kappa - m_0 -\sum_{i\not =0} s^{(-)}_i)d\phi\,,
\eea
at $\theta=0$, while for $\theta= \pi$ one has:
\bea
&&\omega^{(1)}_i={s^{(+)}_i\over 2}\Bigl(-h+{q\over a_i}\Bigr)d\phi\,,\quad \omega^{(3)}_{ij}={s^{(+)}_i s^{(+)}_j\over 2 (a_j -a_i)}d\phi\,, \quad \omega^{(5)}_{ij}=-{s^{(+)}_i s^{(+)}_j\over 2 a_j (a_j -a_i)}d\phi\,,\nonumber\\
&& \omega^{(8)}_{ijk}={s^{(+)}_i s^{(+)}_j s^{(+)}_k\over 2 a_i a_j a_k}d\phi\,,\quad \omega^{(9)}= (\kappa + m_0 +\sum_{i\not =0} s^{(+)}_i)d\phi\,,
\eea
where we have defined
\be
s^{\pm}_i \equiv \mathrm{sign} (r\pm a_i)\,.
\ee
Hence the absence of Dirac-Misner strings imposes the constraints
\bea
&&\sum_{i}  \ell_I   d^{(I)}_i {s^{(-)}_i\over 2}\Bigl(h+{q\over a_i}\Bigr)+ \sum_{i\not = i'} Q^{(I)}_i d^{(I)}_{i'}
{s^{(-)}_i s^{(-)}_{i'}\over 2 (a_{i'}-a_i)}\Bigl(h+{q\over a_{i'}}\Bigr)\nonumber\\
&&\qquad\qquad  + h q \sum_{ijk} d^{(1)}_i d^{(2)}_j d^{(3)}_k {s^{(-)}_i s^{(-)}_{j} s^{(-)}_k\over 2 a_i a_j a_k}+ \kappa - m_0 -\sum_{i\not =0} s^{(-)}_i m_i=0\,,\\
&&-\sum_{i} \ell_I  d^{(I)}_i {s^{(+)}_i\over 2}\Bigl(h-{q\over a_i}\Bigr)+\sum_{i\not = i'} Q^{(I)}_i d^{(I)}_{i'}
{s^{(+)}_i s^{(+)}_{i'}\over 2 (a_{i'}-a_i)}\Bigl(h-{q\over a_{i'}}\Bigr)\nonumber\\
&&\qquad\qquad  + h q \sum_{ijk} d^{(1)}_i d^{(2)}_j d^{(3)}_k {s^{(+)}_i s^{(+)}_{j} s^{(+)}_k\over 2 a_i a_j a_k}+ \kappa + m_0 +\sum_{i\not =0} s^{(+)}_i m_i=0\,.
\eea

Note that, taking into account the possible values of the signs
$s^{(\pm)}_i$, the conditions above imply $N+2$ independent
constraints. One can make these constraints explicit, for example, by
solving them with respect to the $N+2$ variables $\kappa$, $m_0$ and
$m_i$ for $i=1,\ldots,N$.  If one considers, for definiteness, a
configuration in which all the poles $a_i$ lie to the right of the
Taub-NUT center ($0<a_1<\ldots<a_N$), then the regularity constraints
are:

\bea
\kappa&=&-q\sum_{i}{ \ell_I d^{(I)}_i\over 2 a_i}-h \sum_{i\not = i'} {Q^{(I)}_i d^{(I)}_{i'}\over 2 (a_{i'}-a_i)}-h q \sum_{i,j,k} {d^{(1)}_i d^{(2)}_j d^{(3)}_k\over 2 a_i a_j a_k}\,,\label{regularity0}\\
m_0&=& -q\sum_{i}{\ell_I  d^{(I)}_i\over 2 a_i}-h \sum_{i} {Q^{(I)}_0 d^{(I)}_i\over 2 a_i}+q \sum_{i\not = i',i\not=0} {Q^{(I)}_i d^{(I)}_{i'}\over 2 a_{i'} (a_{i'}-a_i)}-h q \sum_{i,j,k} {d^{(1)}_i d^{(2)}_j d^{(3)}_k\over 2 a_i a_j a_k}\,,\label{regularity1}\\
m_i&=& {\ell_I  d^{(I)}_i\over 2} \Bigl(h+{q\over a_i}\Bigr)+\sum_j  {1\over 2 |a_i - a_j|} \Bigl[Q^{(I)}_j d^{(I)}_i \Bigl(h+{q\over a_i}\Bigr)-Q^{(I)}_i d^{(I)}_j \Bigl(h+{q\over a_j}\Bigr)\Bigr]\nonumber\\
&&+{h q\over 2} \Bigl[{d^{(1)}_i d^{(2)}_i d^{(3)}_i \over a_i^3} +{|\epsilon_{IJK}|\over 2} {d^{(I)}_i\over a_i} \sum_{j,k}  \mathrm{sign}(a_j-a_i)  \mathrm{sign}(a_k-a_i){d^{(J)}_j d^{(K)}_k\over a_j a_k} \Bigr]\quad (i\ge1)\,,
\label{regularity2}
\eea
where we have used the convention $\mathrm{sign}(0)=0$.

When there is no black hole and no rotation at the center of Taub-NUT
($Q^{(I)}_0=0$ and $\alpha_0=0$), the metric around $r=0$ is expected
to describe empty space, and hence be completely regular.
As both coordinates $\psi$ and $\phi$
degenerate at $r=0$, regularity requires that $\mu$ and $\omega$
vanish. From (\ref{fullmu}) and (\ref{fullomega}) and the
regularity relations
 (\ref{regularity0}), (\ref{regularity1}) and (\ref{regularity2}), one indeed
finds that $\mu$ and $\omega$ must satisfy:

\bea
&&\mu|_{r=0}=\sum_{i} {\ell_I d^{(I)}_i\over 2 a_i}-\sum_{i\not=i',i\not=0}{Q^{(I)}_i d^{(I)}_{i'}\over 2 a_{i'}(a_{i'}-a_i)}+h \sum_{i,j,k}{d^{(1)}_i d^{(2)}_j d^{(3)}_k\over 2 a_i a_j a_k}+{m_0\over q}=0\,,\nonumber\\
&&\omega|_{r=0}=\Bigl[-\sum_{i} {\ell_I d^{(I)}_i\over 2}\Bigl(h+{q\cos\theta \over a_i}\Bigr)+\sum_{i\not=i',i\not=0}{Q^{(I)}_i d^{(I)}_{i'}\over 2 (a_{i'}-a_i)}\Bigl(h+{q\cos\theta \over a_{i'}}\Bigr)\nonumber\\
&&\quad\qquad  -h q \sum_{i,j,k}{d^{(1)}_i d^{(2)}_j d^{(3)}_k \cos\theta \over 2 a_i a_j a_k}+\kappa -m_0 \cos\theta +\sum_{i\not=0} m_i\Bigr]d\phi=0\,,
\eea
which are automatically implied by (\ref{regularity0}), (\ref{regularity1}) and (\ref{regularity2}). Hence, these relations are enough to guarantee the regularity of the solution at the center of Taub-NUT space.

%%%%%%%%%%%%%%%%%%%%%%%%%%%%%
\subsection{Regularity at the horizons}
%%%%%%%%%%%%%%%%%%%%%%%%%%%%%

It is also important to study the geometry in the vicinity of the poles $\Sigma_i=0$, where, for generic charges and not-too-large angular momenta, we expect to find regular horizons. For this purpose it is convenient to define
\be
I_4 = Z_1  Z_2  Z_3 V - \mu^2 V^2\,.
\ee
The volume element of the horizon around $\Sigma_i=0$ is
\be
\sqrt{g_{H,i}}=\Sigma_i (I_4 \Sigma_i^2 \sin^2\theta_i-\omega_\phi^2)^{1/2}\,.
\ee

Consider first the black hole horizon at $\Sigma_0 \equiv r=0$. The near-horizon expansion gives
\be
I_4\approx {Q_0^{(1)} Q_0^{(2)} Q_0^{(3)} q - \alpha_0^2 \cos^2\theta\over r^4}\,,\quad \omega_\phi \approx \alpha_0 {\sin^2\theta\over r}\,,
\ee
and hence
\be
\sqrt{g_{H,0}}\approx (Q_0^{(1)} Q_0^{(2)} Q_0^{(3)} q -\alpha_0^2)^{1/2} \sin\theta\,.
\ee
Thus we find a horizon of finite   area\footnote{As usual, area means the spatial measure of the three-dimensional horizon of the five-dimensional black hole.} given by:
\be
A_{H,0}= (4\pi q) (4\pi)(Q_0^{(1)} Q_0^{(2)} Q_0^{(3)} q -\alpha_0^2)^{1/2}\,.
\ee
As expected, the black hole at the center is the four-charge rotating black hole constructed
in \cite{BDGRW}, and the parameter $\alpha_0$ encodes its four-dimensional angular momentum.

Consider now the limiting form of the metric near the  $i^{\rm th}$ point (around $\Sigma_i=0$). After several highly non-trivial cancelations one obtains:
\be
I_4=  -2 \alpha_i d_i^{(1)} d_i^{(2)} d_i^{(3)} \Bigl(h+{q\over a_i}\Bigr)^2 {\cos\theta_i\over \Sigma_i^5}+O(\Sigma_i^{-4})
\ee
and
\be
\omega_\phi\sim \Sigma_i^{-1}\,.
\ee
This would lead to closed timelike curves outside the horizon unless the
term of order $\Sigma_i^{-5}$ in $I_4$ vanishes, which
requires\footnote{In the two-center solution of \cite{BDGRW} a non-zero value for
  $\alpha_i$ was required for regularity at the black ring horizon. However, the parameter $\alpha_i$ in \cite{BDGRW} differs from the
  one used here by a constant coming from the gauge choice for  $\mu^{(6)}$, and the two results are consistent.}:
\be
\alpha_i =0\,\quad (i\ge1).
\ee
When this condition is imposed, each point $\Sigma_i=0$ is a black ring horizon of area
\be
A_H = 16 \pi^2 q J_4^{1/2}\,,
\label{area1}
\ee
where $J_4$ is the usual $E_{7(7)}$ quartic invariant that appears in the black ring horizon area \cite{Bena:2004tk}:
\be
J_4 = {1\over 2} \sum_{I<J}  \hat{d}^{(I)}_i \hat{d}^{(J)}_i Q^{(I)}_i Q^{(J)}_i-{1\over 4}\sum_I (\hat{d}^{(I)}_i)^2 (Q^{(I)}_i)^2 - 2 \hat{d}^{(1)}_i \hat{d}^{(2)}_i \hat{d}^{(3)}_i \hat{m}_i\,.
\label{area2}
\ee
In order to bring $J_4$ to its canonical form, we have defined the ``effective'' dipole and angular momentum
parameters;\footnote{The ``effective angular momentum'' that appears in the $J_4$ parameter of the non-BPS
black ring in Taub-NUT constructed in \cite{BDGRW} is not $\hat{m}_i$
but
\be
 \hat  m_i^\mathrm{BDGRW}  \equiv   \hat {m}_i - {q\over 2 a_i^2} \Bigl(h+{q\over a_i}\Bigr)^{-2} Q_0^{(I)} \hat{d}^{(I)}_i \qquad \Leftrightarrow \qquad m_i^\mathrm{here}= m_i^\mathrm{BDGRW} + {q\over 2 a_i^2} Q^{(I)}_0 d^{(I)}_i\,.
\ee
We find here, instead, that $J_4$ simply depends on $\hat{m}_i$. The two results are consistent
because here we are using a different (and more natural) gauge choice originating from a different
definition of $\mu^{(5)}_{0i}$ and reflected in the equation for $m_i^\mathrm{here}$. }
\be
\hat{d}^{(I)}_i=  \Bigl(h+{q\over a_i}\Bigr) d^{(I)}_i\,,\quad \hat{m}_i= \Bigl(h+{q\over a_i}\Bigr)^{-1}m_i \,.
\ee
Note that the result (\ref{area1}) and (\ref{area2}) coincides with the one for an isolated BPS black ring carrying charges $\hat{d}^{(I)}_i$, $Q^{I}_i$ and $\hat{m}_i$: the area of the $i^{\rm th}$ horizon is not affected by the
presence of the other horizons nor by the switch of orientation of the base space that is characteristic of our non-BPS solutions.

If one chooses units such that the five-dimensional Newton's constant is $G_5={\pi\over 4}$ and the three tori
have equal sizes, the integer M2, M5 and KK momentum charges carried by the $i^{\rm th}$  center are:
\be
n^{(I)}_i=-{\hat{d}^{(I)}_i\over 2}\,,\quad N^{(I)}_i = {Q^{(I)}_i\over 4}\,,\quad J^{(KK)}_i=-{\hat{m}_i\over 8}\,.
\ee

One can also construct solutions in which some of the centers do not have three M2 charges and three M5
charges, but only two M2 charges and one M5 charge. These solutions describe now two-charge round supertubes \cite{supertube}, and the geometry near an individual supertube is expected to be smooth in the duality frame in which the dipole charge of the tube corresponds to KK-monopoles, and the electric charges correspond to D1 and D5 brane \cite{Mathur-Lunin,LMM,our-supertube}).

For the supertube with dipole charge corresponding to, say, $K^3$, this regularity condition is \cite{our-supertube}:
\be
\lim_{\Sigma_i\to 0}\Sigma_i^2 (Z_3  V (K^{(3)})^2 -2 \mu V K^{(3)}+ Z_1 Z_2)=0\,.
\label{supertuberegularity}
\ee
Just as for black rings, this requires that the ``dipole'' harmonic term in $M$ vanish (otherwise $\mu V K_i^3\sim \Sigma_i^{-3}$):
\be
\alpha_i=0\,.
\ee
Furthermore, equation (\ref{supertuberegularity}) implies the usual supertube regularity condition:
\be
2 d^3_i m_i =  Q^2_i Q^1_i\,.
\ee

%%%%%%%%%%%%%%%%%%%%%%%%%%%%%
\subsection{Scaling solutions}
%%%%%%%%%%%%%%%%%%%%%%%%%%%%%

Consider the limit in which the positions of the centers are scaled to zero ($a_i\ll {q\over h}$). In this limit the regularity conditions (\ref{regularity1}) and (\ref{regularity2}), when written in terms of the quantized charge parameters  $\hat{d}^{(I)}_i$, $Q^{I}_i$ and $\hat{m}_i$, reduce to:
\bea
m_0&=& -\sum_{i}{\ell_I \hat{d}^{(I)}_i\over 2}-{h\over q} \sum_{i} {Q^{(I)}_0 \hat{d}^{(I)}_i\over 2}+ \sum_{i\not = i',i\not=0} {Q^{(I)}_i \hat{d}^{(I)}_{i'}\over 2 (a_{i'}-a_i)}-{h\over q^2} \sum_{i,j,k} {\hat{d}^{(1)}_i \hat{d}^{(2)}_j \hat{d}^{(3)}_k\over 2}\,,\label{regularitybis1}\\
q{\hat{m}_i\over a_i}&=& {\ell_I  \hat{d}^{(I)}_i\over 2} +\sum_j  {1\over 2 |a_i - a_j|} \Bigl[Q^{(I)}_j \hat{d}^{(I)}_i -Q^{(I)}_i \hat{d}^{(I)}_j\Bigr]\nonumber\\
&&+{h\over 2 q^2} \Bigl[\hat{d}^{(1)}_i \hat{d}^{(2)}_i \hat{d}^{(3)}_i  +{|\epsilon_{IJK}|\over 2} \hat{d}^{(I)}_i \sum_{j,k}  \mathrm{sign}(a_j-a_i)  \mathrm{sign}(a_k-a_i)\hat{d}^{(J)}_j \hat{d}^{(K)}_k\Bigr]\quad (i\ge1)\,.
\label{regularitybis2}
\eea
These equations are now linear in the inverse of the inter-center distance, much as they are for BPS solutions.

As the parameters $\hat{d}^{(I)}_i$, $Q^{I}_i$ and $\hat{m}_i$ with
$i>0$ are associated to quantized charges, their value is to be kept
finite while the $a_i$'s are scaled to zero. Note however that $m_0$ does
not correspond to any quantized charge, but is a parameter needed for
regularity, as indicated by (\ref{regularity0}).
Hence, one should think about equation (\ref{regularitybis1}) (or (\ref{regularity1}) in the full solution) as determining the value of a parameter of the solution as a function of the charges and the positions of the centers, and about equations  (\ref{regularitybis2}) (or  (\ref{regularitybis2}) in the full solution) as the ``bubble equations'' of the system, that determine the inter-center distances as a function of the charges and the moduli.

In the small $a_i$ limit, the non-BPS bubble equations become:
\bea
 \sum_j  {1\over 2 |a_i - a_j|} \Bigl[Q^{(I)}_j \hat{d}^{(I)}_i -Q^{(I)}_i \hat{d}^{(I)}_j\Bigr]=q{\hat{m}_i\over a_i}\,,
\eea
which coincides with the scaling limit of the BPS bubble equations.

%%%%%%%%%%%%%%%%%%%%%%%%%%%%%%%%%%%%%
\section{Conclusions and future directions}
%%%%%%%%%%%%%%%%%%%%%%%%%%%%%%%%%%%%%

We have constructed almost-BPS multi-center solutions that describe a
black hole and an arbitrary number of black rings in Taub-NUT. This
solution descends to four dimensions to a multi-center configurations
containing one rotating $\overline{\rm D6}$-(D2)$^3$ and an arbitrary
number of collinear (D4)$^3$-(D2)$^3$-D0 black holes. These solutions admit
scaling regimes where some, or all, of the centers get very close to
each other (in $\IR^3$ coordinate distance), and the throats of the
black holes that are scaling join into a bigger throat. Furthermore,
since the bubble equations are insensitive to the four-dimensional
rotation of the black hole, we can obtain scaling solutions that have
a non-zero four-dimensional angular momentum.

There are several obvious  directions for future research.  On a technical level, it should be possible to
generalize our results to ``tilted'' black rings where the centers are not co-linear in the $\IR^3$ base.
Preliminary calculations suggest that while this should in principle be possible, it is technically quite
complicated.

On a more fundamental and physical level it would be interesting  to determine whether
the non-BPS bubble equations can be derived from a microscopic quiver
perspective in the way the BPS ones were derived in \cite{denef}.
Given the complicated structure of the bubble equations, and the fact
that they do not depend on the four-dimensional angular momentum, this
would be quite spectacular. It would also hint at the existence of
non-renormalization theorems that apply both to BPS and to non-BPS
multi-center solutions, and may allow a moduli-space quantization of
these solutions similar to that of \cite{sheer}.

It is also interesting to explore the lines of marginal stability in
the moduli space of the almost-BPS solutions. Note that, unlike their
BPS counterparts, these solutions are completely independent of some of
the moduli: for example, Wilson lines along the
Taub-NUT fiber at infinity correspond to adding constants to the $K^{(I)}$
harmonic functions, which does not affect at all the metric or the
Maxwell fields.

It is equally important to try to use our multi-center almost-BPS
solutions to construct smooth horizonless black hole microstate
geometries corresponding to microstates of rotating non-BPS
four-dimensional black holes. This is however not as straightforward
as for BPS solutions. Indeed, if one considers an almost-BPS solution with a
multi-center Gibbons-Hawking or Taub-NUT base, the flux on a two-cycle running
between two centers is anti-self-dual, and hence non-normalizable.
Such solutions thus tend to be unphysical and so such fluxes should be set to zero.
If one then builds solutions with multiple D6 centers but without fluxes, these
centers are always mutually local (there is no arrow between them in
the quiver description), and the solution one builds is uninteresting.

One way to proceed is to relax at first the requirement of
smoothness, and to focus rather on ``primitive'' centers (that
correspond to fluxed D4, fluxed D2, or D0 branes, and that preserve
locally 16 supercharges). Then our solutions contain, for example, a
four-point quiver that has one D6 and three mutually-nonlocal fluxed
D4 centers (which can also be thought of as supertubes that have three different
kinds of dipole charges). This quiver has arrows running between all
centers, admits scaling solutions, and can have overall charges
corresponding to a rotating black hole of macroscopic horizon area.
Furthermore, one can argue that upon a chain of
dualities\footnote{For example three spectral flow transformations of the type
  explored in \cite{SpecFlow}} this solution can be brought to a
duality frame in which it is completely smooth, much like a fluxed D4 
(also known as a supertube) sources a smooth supergravity solution in
the duality frame where the supertube has KKM dipole charge
\cite{Mathur-Lunin,LMM,our-supertube}. We believe this route will
yield rather generic smooth microstates of non-BPS extremal black
holes, and will help extend the fuzzball proposal \cite{fuzzball-reviews} beyond
supersymmetric settings.

Last, but not least it is important to understand the other
circumstances in which Einstein's equations can factorize and one has
the hope of constructing solutions systematically. As we will see in a
forthcoming publication \cite{nextpaper}, the almost-BPS solutions are
not the most generic non-BPS solutions that factorize, and there may
exist routes to classify and find all the extremal multi-center
non-BPS solutions one can build in four or five-dimensional
supergravity.

%%%%%%%%%%%%%%%%%%%%%%%%%%%%%%%%%%%%%
\bigskip
\leftline{\bf Acknowledgments}
\smallskip
We would like to thank Gianguido Dall'Agata for interesting discussions.
NPW is very grateful to the Mitchell Family Foundation for hospitality at the Cook's Branch Conservancy and to the IPhT(SPhT), CEA-Saclay for hospitality
while this work was in various stages of development.
The work of IB, CR and SG was supported in part by the
DSM CEA-Saclay, by the ANR grants BLAN 06-3-137168 and 08-JCJC-0001-01
and by the Marie Curie IRG 046430.
The work of NPW was supported in part by DOE grant DE-FG03-84ER-40168.

%%%%%%%%%%%%%%%%%%%%%%%%%%%%%%%%%%%%%

%%%%%%%%%%%%%%%%%%%%%%%%%%%%%%%%%%%


\begin{thebibliography}{99}
%%%%%%%%%%%%%%%%%%%%%%%%%%%%%%%%%%%%%

\bibitem{denef}
 F.~Denef,
  ``Supergravity flows and D-brane stability,''
  JHEP {\bf 0008}, 050 (2000)
  [arXiv:hep-th/0005049].

  F.~Denef,
  ``Quantum quivers and Hall/hole halos,''
  JHEP {\bf 0210}, 023 (2002)
  [arXiv:hep-th/0206072].

  B.~Bates and F.~Denef,
  ``Exact solutions for supersymmetric stationary black hole composites,''
  [arXiv:hep-th/0304094].



 %\cite{Gauntlett:2002nw}
\bibitem{Gauntlett:2002nw}
  J.~P.~Gauntlett, J.~B.~Gutowski, C.~M.~Hull, S.~Pakis and H.~S.~Reall,
``All supersymmetric solutions of minimal supergravity in five dimensions,''
  Class.\ Quant.\ Grav.\  {\bf 20}, 4587 (2003)
  [arXiv:hep-th/0209114].
  %%CITATION = HEP-TH 0209114;%%


%\cite{Gauntlett:2004qy}
\bibitem{Gauntlett:2004qy}
  J.~P.~Gauntlett and J.~B.~Gutowski,
``General concentric black rings,''
  Phys.\ Rev.\ D {\bf 71}, 045002 (2005)
  [arXiv:hep-th/0408122].
  %%CITATION = HEP-TH 0408122;%%


%\cite{Bena:2005va}
\bibitem{Bena:2005va}
  I.~Bena and N.~P.~Warner,
 ``Bubbling supertubes and foaming black holes,''
  Phys.\ Rev.\  D {\bf 74}, 066001 (2006)
  [arXiv:hep-th/0505166].
  %%CITATION = PHRVA,D74,066001;%%

%\cite{Berglund:2005vb}
\bibitem{Berglund:2005vb}
  P.~Berglund, E.~G.~Gimon and T.~S.~Levi,
  ``Supergravity microstates for BPS black holes and black rings,''
  JHEP {\bf 0606}, 007 (2006)
  [arXiv:hep-th/0505167].

%\cite{Saxena:2005uk}
\bibitem{Saxena:2005uk}
  A.~Saxena, G.~Potvin, S.~Giusto and A.~W.~Peet,
  ``Smooth geometries with four charges in four dimensions,''
  JHEP {\bf 0604} (2006) 010
  [arXiv:hep-th/0509214].
  %%CITATION = JHEPA,0604,010;%%



\bibitem{4D5Dring}
%\cite{Elvang:2005sa}
  H.~Elvang, R.~Emparan, D.~Mateos and H.~S.~Reall,
  ``Supersymmetric 4D rotating black holes from 5D black rings,''
  JHEP {\bf 0508}, 042 (2005)
  [arXiv:hep-th/0504125].
  %%CITATION = JHEPA,0508,042;%%

%\cite{Gaiotto:2005xt}
%\bibitem{Gaiotto:2005xt}
  D.~Gaiotto, A.~Strominger and X.~Yin,
  ``5D black rings and 4D black holes,''
  JHEP {\bf 0602} (2006) 023
  [arXiv:hep-th/0504126].
  %%CITATION = JHEPA,0602,023;%%

%\cite{Bena:2005ni}
  I.~Bena, P.~Kraus and N.~P.~Warner,
  ``Black rings in Taub-NUT,''
  Phys.\ Rev.\  D {\bf 72}, 084019 (2005)
  [arXiv:hep-th/0504142].


\bibitem{denef-moore}
%\cite{Denef:2007vg}
%\bibitem{Denef:2007vg}
  F.~Denef and G.~W.~Moore,
  ``Split states, entropy enigmas, holes and halos,''
  arXiv:hep-th/0702146.
  %%CITATION = HEP-TH/0702146;%%


\bibitem{OSV}
%\cite{Ooguri:2004zv}
%\bibitem{Ooguri:2004zv}
  H.~Ooguri, A.~Strominger and C.~Vafa,
  ``Black hole attractors and the topological string,''
  Phys.\ Rev.\  D {\bf 70}, 106007 (2004)
  [arXiv:hep-th/0405146].
  %%CITATION = PHRVA,D70,106007;%%


\bibitem{us}
%\cite{Bena:2006kb}
%\bibitem{Bena:2006kb}
  I.~Bena, C.~W.~Wang and N.~P.~Warner,
  ``Mergers and Typical Black Hole Microstates,''
  JHEP {\bf 0611}, 042 (2006)
  [arXiv:hep-th/0608217].
  %%CITATION = JHEPA,0611,042;%%

%\cite{Bena:2007qc}
\bibitem{Bena:2007qc}
  I.~Bena, C.~W.~Wang and N.~P.~Warner,
  ``Plumbing the Abyss: Black Ring Microstates,''
  JHEP {\bf 0807} (2008) 019
  [arXiv:0706.3786 [hep-th]].
  %%CITATION = JHEPA,0807,019;%%


\bibitem{stability} For a few examples see:

  A.~Sen,
  ``Walls of Marginal Stability and Dyon Spectrum in N=4 Supersymmetric
  String Theories,''
  JHEP {\bf 0705}, 039 (2007)
  [arXiv:hep-th/0702141].
  %%CITATION = JHEPA,0705,039;%%

%\cite{Gaiotto:2009hg}
%\bibitem{Gaiotto:2009hg}
  D.~Gaiotto, G.~W.~Moore and A.~Neitzke,
  ``Wall-crossing, Hitchin Systems, and the WKB Approximation,''
  arXiv:0907.3987 [hep-th].
  %%CITATION = ARXIV:0907.3987;%%


%\cite{Manschot:2009ia}
%\bibitem{Manschot:2009ia}
  J.~Manschot,
  ``Stability and duality in N=2 supergravity,''
  arXiv:0906.1767 [hep-th].
  %%CITATION = ARXIV:0906.1767;%%

\bibitem{sheer}
%\cite{deBoer:2008zn}
%\bibitem{deBoer:2008zn}
  J.~de Boer, S.~El-Showk, I.~Messamah and D.~V.~d.~Bleeken,
  ``Quantizing N=2 multi-center Solutions,''
  arXiv:0807.4556 [hep-th].
  %%CITATION = ARXIV:0807.4556;%%
  
%\cite{deBoer:2009un}
%\bibitem{deBoer:2009un}
  J.~de Boer, S.~El-Showk, I.~Messamah and D.~V.~d.~Bleeken,
  ``A bound on the entropy of supergravity?,''
  arXiv:0906.0011 [hep-th].
  %%CITATION = ARXIV:0906.0011;%%


%\cite{Bena:2004de}
\bibitem{Bena:2004de}
  I.~Bena and N.P.~Warner,
``One ring to rule them all ... and in the darkness bind them?,''
  Adv.\ Theor.\ Math.\ Phys.\  {\bf 9}, 667 (2005)
  [arXiv:hep-th/0408106].
  %%CITATION = 00203,9,667;%%

%\cite{Gutowski:2004yv}
\bibitem{Gutowski:2004yv}
  J.~B.~Gutowski and H.~S.~Reall,
  ``General supersymmetric AdS(5) black holes,''
  JHEP {\bf 0404}, 048 (2004)
  [arXiv:hep-th/0401129].
  %%CITATION = JHEPA,0404,048;%%

\bibitem{JMaRT-all}

%\cite{Jejjala:2005yu}
%\bibitem{Jejjala:2005yu}
  V.~Jejjala, O.~Madden, S.~F.~Ross and G.~Titchener,
  ``Non-supersymmetric smooth geometries and D1-D5-P bound states,''
  Phys.\ Rev.\  D {\bf 71}, 124030 (2005)
  [arXiv:hep-th/0504181].
  %%CITATION = PHRVA,D71,124030;%%

%Giusto:2007tt,Giusto:2007fx,Ford:2007th,Camps:2008hb,JMaRT-flow}
%\cite{Giusto:2007tt}
%\bibitem{Giusto:2007tt}
  S.~Giusto, S.~F.~Ross and A.~Saxena,
  ``Non-supersymmetric microstates of the D1-D5-KK system,''
  JHEP {\bf 0712}, 065 (2007)
  [arXiv:0708.3845 [hep-th]].
  %%CITATION = JHEPA,0712,065;%%

%\cite{Camps:2008hb}
%\bibitem{Camps:2008hb}
  J.~Camps, R.~Emparan, P.~Figueras, S.~Giusto and A.~Saxena,
  ``Black Rings in Taub-NUT and D0-D6 interactions,''
  JHEP {\bf 0902}, 021 (2009)
  [arXiv:0811.2088 [hep-th]].
  %%CITATION = JHEPA,0902,021;%%

%\cite{Giusto:2007fx}
%\bibitem{Giusto:2007fx}
  S.~Giusto and A.~Saxena,
  ``Stationary axisymmetric solutions of five dimensional gravity,''
  Class.\ Quant.\ Grav.\  {\bf 24}, 4269 (2007)
  [arXiv:0705.4484 [hep-th]].
  %%CITATION = CQGRD,24,4269;%%

%\cite{Ford:2007th}
%\bibitem{Ford:2007th}
  J.~Ford, S.~Giusto, A.~Peet and A.~Saxena,
  ``Reduction without reduction: Adding KK-monopoles to five dimensional
  stationary axisymmetric solutions,''
  Class.\ Quant.\ Grav.\  {\bf 25}, 075014 (2008)
  [arXiv:0708.3823 [hep-th]].
  %%CITATION = CQGRD,25,075014;%%


%\bibitem{JMaRT-flow}
%\cite{AlAlawi:2009qe}
  J.~H.~Al-Alawi and S.~F.~Ross,
  ``Spectral Flow of the Non-Supersymmetric Microstates of the D1-D5-KK
  System,''
  arXiv:0908.0417 [hep-th].
  %%CITATION = ARXIV:0908.0417;%%


%\cite{Gaiotto:2007ag}
\bibitem{Gaiotto-Finn}
  D.~Gaiotto, W.~W.~Li and M.~Padi,
  ``Non-Supersymmetric Attractor Flow in Symmetric Spaces,''
  JHEP {\bf 0712}, 093 (2007)
  [arXiv:0710.1638 [hep-th]].
  %%CITATION = JHEPA,0712,093;%%
  %\cite{Gimon:2009gk}

%\bibitem{Gimon:2009gk}
  E.~G.~Gimon, F.~Larsen and J.~Simon,
  ``Constituent Model of Extremal non-BPS Black Holes,''
  JHEP {\bf 0907}, 052 (2009)
  [arXiv:0903.0719 [hep-th]].
  %%CITATION = JHEPA,0907,052;%%

  

%\cite{Ceresole:2007wx}
\bibitem{Ceresole:2007wx}
  A.~Ceresole and G.~Dall'Agata,
  ``Flow Equations for Non-BPS Extremal Black Holes,''
  JHEP {\bf 0703}, 110 (2007)
  [arXiv:hep-th/0702088].
  %%CITATION = JHEPA,0703,110;%%

%\cite{Lopes Cardoso:2007ky}
\bibitem{factorization}
  G.~Lopes Cardoso, A.~Ceresole, G.~Dall'Agata, J.~M.~Oberreuter and J.~Perz,
  ``First-order flow equations for extremal black holes in very special
  geometry,''
  JHEP {\bf 0710}, 063 (2007)
  [arXiv:0706.3373 [hep-th]].
  %%CITATION = JHEPA,0710,063;%%


%\cite{Andrianopoli:2007gt}
%\bibitem{Andrianopoli:2007gt}
  L.~Andrianopoli, R.~D'Auria, E.~Orazi and M.~Trigiante,
  ``First Order Description of Black Holes in Moduli Space,''
  JHEP {\bf 0711}, 032 (2007)
  [arXiv:0706.0712 [hep-th]].
  %%CITATION = JHEPA,0711,032;%%

%\cite{Ferrara:2008ap}
%\bibitem{Ferrara:2008ap}
  S.~Ferrara, A.~Gnecchi and A.~Marrani,
  ``d=4 Attractors, Effective Horizon Radius and Fake Supergravity,''
  Phys.\ Rev.\  D {\bf 78}, 065003 (2008)
  [arXiv:0806.3196 [hep-th]].
  %%CITATION = PHRVA,D78,065003;%%

%\cite{Bellucci:2008sv}
%\bibitem{Bellucci:2008sv}
  S.~Bellucci, S.~Ferrara, A.~Marrani and A.~Yeranyan,
  ``stu Black Holes Unveiled,''
  arXiv:0807.3503 [hep-th].
  %%CITATION = ARXIV:0807.3503;%%


%\cite{Goldstein:2008fq}
\bibitem{Goldstein}
  K.~Goldstein and S.~Katmadas,
  ``Almost BPS black holes,''
  arXiv:0812.4183 [hep-th].
  %%CITATION = ARXIV:0812.4183;%%

\bibitem{BDGRW}
%\cite{Bena:2009ev}
%\bibitem{Bena:2009ev}
  I.~Bena, G.~Dall'Agata, S.~Giusto, C.~Ruef and N.~P.~Warner,
  ``Non-BPS Black Rings and Black Holes in Taub-NUT,''
  JHEP {\bf 0906}, 015 (2009)
  [arXiv:0902.4526 [hep-th]].
  %%CITATION = JHEPA,0906,015;%%


\bibitem{emparan-horowitz}
%\cite{Emparan:2006it}
%\bibitem{Emparan:2006it}
  R.~Emparan and G.~T.~Horowitz,
  ``Microstates of a neutral black hole in M theory,''
  Phys.\ Rev.\ Lett.\  {\bf 97}, 141601 (2006)
  [arXiv:hep-th/0607023].
  %%CITATION = PRLTA,97,141601;%%


\bibitem{gg1} Gauntlett Gutowski 1
%\cite{Gauntlett:2004wh}
%\bibitem{Gauntlett:2004wh}
  J.~P.~Gauntlett and J.~B.~Gutowski,
  ``Concentric black rings,''
  Phys.\ Rev.\  D {\bf 71}, 025013 (2005)
  [arXiv:hep-th/0408010].
  %%CITATION = PHRVA,D71,025013;%%


\bibitem{SpecFlow} Spectral Flow
%\cite{Bena:2008wt}
%\bibitem{Bena:2008wt}
  I.~Bena, N.~Bobev and N.~P.~Warner,
 ``Spectral Flow, and the Spectrum of Multi-Center Solutions,''
  Phys.\ Rev.\  D {\bf 77}, 125025 (2008)
  [arXiv:0803.1203 [hep-th]].
  %%CITATION = PHRVA,D77,125025;%%


%\cite{Bena:2004tk}
\bibitem{Bena:2004tk}
  I.~Bena and P.~Kraus,
  ``Microscopic description of black rings in AdS/CFT,''
  JHEP {\bf 0412}, 070 (2004)
  [arXiv:hep-th/0408186].
  %%CITATION = JHEPA,0412,070;%%



\bibitem{supertube}
%\cite{Mateos:2001qs}
  D.~Mateos and P.~K.~Townsend,
  ``Supertubes,''
  Phys.\ Rev.\ Lett.\  {\bf 87}, 011602 (2001)
  [arXiv:hep-th/0103030].
%\cite{Emparan:2001ux}
  R.~Emparan, D.~Mateos and P.~K.~Townsend,
  ``Supergravity supertubes,''
  JHEP {\bf 0107}, 011 (2001)
  [arXiv:hep-th/0106012].


\bibitem{Mathur-Lunin}
%\cite{Lunin:2001fv}
%\bibitem{Lunin:2001fv}
  O.~Lunin and S.~D.~Mathur,
  ``Metric of the multiply wound rotating string,''
  Nucl.\ Phys.\  B {\bf 610}, 49 (2001)
  [arXiv:hep-th/0105136].
  %%CITATION = NUPHA,B610,49;%%


\bibitem{LMM}
%\cite{Lunin:2002iz}
%\bibitem{Lunin:2002iz}
  O.~Lunin, J.~M.~Maldacena and L.~Maoz,
  ``Gravity solutions for the D1-D5 system with angular momentum,''
  arXiv:hep-th/0212210.
  %%CITATION = HEP-TH/0212210;%%



\bibitem{our-supertube}
%\cite{Bena:2008dw}
%\bibitem{Bena:2008dw}
  I.~Bena, N.~Bobev, C.~Ruef and N.~P.~Warner,
  ``Supertubes in Bubbling Backgrounds: Born-Infeld Meets Supergravity,''
  arXiv:0812.2942 [hep-th].
  %%CITATION = ARXIV:0812.2942;%%


\bibitem{fuzzball-reviews}
S.~D.~Mathur,
  ``The fuzzball proposal for black holes: An elementary review,''
    Fortsch.\ Phys.\  {\bf 53}, 793 (2005)
%[arXiv:hep-th/0502050].

 %\cite{Bena:2007kg}
% \bibitem{Bena:2007kg}
  I.~Bena and N.~P.~Warner,
``Black holes, black rings and their microstates,''
  arXiv:hep-th/0701216.
  %%CITATION = HEP-TH/0701216;%%
Lect.\ Notes Phys.\  {\bf 755}, 1 (2008)



%\bibitem{kostas}
  K.~Skenderis and M.~Taylor,
  ``The fuzzball proposal for black holes,''
  Phys.\ Rept.\  {\bf 467}, 117 (2008)
  %%CITATION = PRPLC,467,117;%%

%\bibitem{jan}
 V.~Balasubramanian, J.~de Boer, S.~El-Showk and I.~Messamah,
  ``Black Holes as Effective Geometries,''
  Class.\ Quant.\ Grav.\  {\bf 25} (2008) 214004
  [arXiv:0811.0263 [hep-th]].
  %%CITATION = CQGRD,25,214004;%%


\bibitem{nextpaper}
I.~Bena, S.~Giusto, C.~Ruef and N.~P.~Warner, to appear






%%%%%%%%%%%%%%%%%%%%%%%%%%%%%%%%%%%%%
\end{thebibliography}
\end{document}